\documentclass[5p]{elsarticle}
\usepackage[hyperindex,breaklinks]{hyperref}
\hypersetup{
   colorlinks=true,%
   citecolor=blue,%
   filecolor=blue,%
   linkcolor=blue,%
   urlcolor=blue
}

\newcommand{\titleinfo}{A search for high-mass resonances decaying to $\tau^{+}\tau^{-}$ in $pp$ collisions at $\sqrt{s}=7\TeV$ with the ATLAS detector}
\newcommand{\abstractinfo}{
This Letter presents a search for high-mass resonances decaying into \tptm{}
final states using proton-proton collisions at $\sqrt{s}= 7\TeV$ produced by
the Large Hadron Collider. The data were recorded with the ATLAS detector and
correspond to an integrated luminosity of 4.6~\ifb{}.  No statistically 
significant excess above the Standard Model expectation is observed; 95\% 
credibility upper limits are set on the cross section times branching fraction 
of $Z'$ resonances decaying into \tptm{} pairs as a function of the resonance 
mass.  As a result, $Z'$ bosons of the Sequential Standard Model with masses 
less than 1.40~TeV are excluded at 95\% credibility.
}

\usepackage{preprintcover}  
\PreprintCoverPaperTitle{\titleinfo}  
\PreprintIdNumber{CERN-PH-EP-2012-280}  
\PreprintCoverAbstract{\abstractinfo}  
\PreprintJournalName{Phys. Lett. B}  

\def\met{\ensuremath{E_{\mathrm{T}}^{\mathrm{miss}}}} 
\def\TeV{\ifmmode {\mathrm{\ Te\kern -0.1em V}}\else
                   \textrm{Te\kern -0.1em V}\fi}%
\def\GeV{\ifmmode {\mathrm{\ Ge\kern -0.1em V}}\else
                   \textrm{Ge\kern -0.1em V}\fi}%

\let\gev=\GeV
\def\ifb{\mbox{fb$^{-1}$}}
\def\pt{\ensuremath{p_{\mathrm{T}}}} 
\def\et{\ensuremath{E_{\mathrm{T}}}} 

\newcommand{\pzetavis}{\ensuremath{p^\mathrm{vis}_{\zeta}}}

\newcommand{\tptm}{\ensuremath{\tau^{+}\tau^{-}}}
\newcommand{\tauh}{\ensuremath{\tau_{\rm had}}}
\mathchardef\mhyphen="2D
\newcommand{\tauhvis}{\ensuremath{\tau_{\rm had \mhyphen vis}}}
\newcommand{\taue}{\ensuremath{\tau_{e}}}
\newcommand{\taumu}{\ensuremath{\tau_{\mu}}}
\newcommand{\taul}{\ensuremath{\tau_{\rm lep}}}

\newcommand{\hadhad}{\ensuremath{\tauh\tauh}}
\newcommand{\muhad}{\ensuremath{\taumu\tauh}}
\newcommand{\ehad}{\ensuremath{\taue\tauh}}
\newcommand{\lephad}{\ensuremath{\taul\tauh}}
\newcommand{\emu}{\ensuremath{\taue\taumu}}

\newcommand{\Zprime}{\ensuremath{Z^{\prime}}}
\newcommand{\ZprimeSSM}{\ensuremath{\Zprime_\mathrm{SSM}}}
\newcommand{\DY}{\ensuremath{Z/\gamma^{\ast}\to\tau\tau}}
\newcommand{\DYonly}{\ensuremath{Z/\gamma^{\ast}}}
\newcommand{\DYjets}{\ensuremath{Z/\gamma^{\ast}{\rm +jets}}}
\newcommand{\Ztautau}{\ensuremath{Z\to\tau\tau}}

\newcommand{\Zmumujets}{\ensuremath{Z(\to\mu\mu){\rm+jets}}}
\newcommand{\Zlljets}{\ensuremath{Z(\to\ell\ell){\rm+jets}}}

\newcommand{\ZP}{\ensuremath{Z'\to\tau\tau}}

\newcommand{\ZPSSM}{\ensuremath{Z'_{\rm SSM}}}
\newcommand{\Wtaunu}{\ensuremath{W\to\tau\nu}}
\newcommand{\Wtaunujets}{\ensuremath{W(\to\tau\nu){\rm+jets}}}
\newcommand{\Wjets}{\ensuremath{W{\rm+jets}}}
\newcommand{\Wlnujets}{\ensuremath{W(\to\ell\nu){\rm+jets}}}

\newcommand{\Wgammajets}{\ensuremath{W\gamma}{\rm+jets}}
\newcommand{\multijet}{multijet}

\newcommand{\singletop}{single top-quark}

\newcommand{\diboson}{diboson}
\newcommand{\Diboson}{Diboson}
\newcommand{\ttbar}{\ensuremath{t\bar{t}}}

\newcommand{\ditau}{ditau}
\newcommand{\diele}{dielectron}
\newcommand{\dimuon}{dimuon}

\newcommand{\metscalar}{\ensuremath{\met}}

\newcommand{\mt}{\ensuremath{m_{\rm T}}}
\newcommand{\mT}{\ensuremath{m_{\rm T}}}
\newcommand{\MT}{\ensuremath{m_{\rm T}^{\rm tot}}}

\newcommand{\Niso}{\mbox{\ensuremath{ N^{\mathrm{iso}} }}}
\newcommand{\Nantiiso}{\ensuremath{ N^{\mathrm{anti}\mbox{--}\mathrm{iso}} }}

\newcommand{\ptone}{\ensuremath{p_{\rm T1}}}
\newcommand{\pttwo}{\ensuremath{p_{\rm T2}}}

\newcommand{\Conemiss}{\ensuremath{C_{1}}}
\newcommand{\Ctwomiss}{\ensuremath{C_{2}}}
\newcommand{\Conetwo}{\ensuremath{C}}

\title{\titleinfo}

\begin{document}

\begin{frontmatter}

\author{The ATLAS Collaboration}

\begin{abstract}
\abstractinfo
\end{abstract}


\end{frontmatter}


\section{Introduction}
\label{sec:introduction}

Many extensions of the Standard Model (SM), motivated by grand unification,
predict additional heavy gauge bosons
~\cite{Langacker:2008,Hewett:1989,Cvetic:1995zs,Leike:1998wr,Rizzo:2006nw,Diener:2010sy}.
As lepton universality is not necessarily a requirement for these new gauge
bosons, it is essential to search in all decay modes.  In particular, some
models with extended weak or hypercharge gauge groups that offer an
explanation for the high mass of the top quark predict that such bosons
preferentially couple to third-generation fermions~\cite{ZprimeToThirdGen}.

This Letter presents the first search for high-mass resonances decaying into
$\tau^{+}\tau^{-}$ pairs using the ATLAS detector~\cite{AtlasDetectorPaper}.
The Sequential Standard Model (SSM) is a benchmark model that contains a heavy
neutral gauge boson, \ZprimeSSM{}, with the same couplings to fermions as the
$Z$ boson of the SM.  This model is used to optimise the event selection of the
search; limits on the cross section times $\tptm$ branching fraction of a 
generic neutral resonance are reported. 

\begin{sloppypar}
Direct searches for high-mass \ditau{} resonances have been performed
previously by the CDF~\cite{CDFZprimeTauTau:2005} and CMS~\cite{CMSZprime5fb}
collaborations. The latter search sets the most stringent 95\% confidence level
limits and excludes \ZprimeSSM{} masses below $1.4\TeV$, with an expected limit
of $1.1\TeV$, using 4.9~\ifb{} of integrated luminosity at $\sqrt{s}=7\TeV$.
Indirect limits on $Z'$ bosons with non-universal flavour couplings have been
set using measurements from LEP and LEP~II~\cite{ZprimeMassLimitsFromLEP} and
translate to a lower bound on the $Z'$ mass of $1.09\TeV$.  For comparison, the
most stringent limits on \ZprimeSSM{} in the \diele{} and \dimuon{} decay
channels combined are $2.2\TeV$ from ATLAS~\cite{ATLAS-CONF-2012-007} and
$2.3\TeV$ from CMS~\cite{CMS-PAS-EXO-11-019}.
\end{sloppypar}

Tau leptons can decay into a charged lepton and two neutrinos ($\taul = \taue$
or \taumu{}), or hadronically (\tauh{}), predominantly into one or three
charged pions, a neutrino and often additional neutral pions. The \hadhad{}
(branching ratio, BR=42\%), \muhad{} (BR=23\%), \ehad{} (BR=23\%) and \emu{}
(BR=6\%) channels are analysed.  Due to the different dominant background contributions and
signal sensitivities, each channel is analysed separately and a statistical
combination is used to maximise the sensitivity.

While the expected natural width of the \ZprimeSSM{} is small, approximately
3\% of the \Zprime{} mass, the mass resolution is 30--50\% in $\tptm$ decay
modes due to the undetected neutrinos from the tau decays.  Therefore, a
counting experiment is performed in all analysis channels from events that pass
a high-mass requirement.

\section{Event samples}
\label{sec:atlas_detector_and_samples}

The data used in this search were recorded with the ATLAS detector in
proton-proton ($pp$) collisions at a centre-of-mass energy of $\sqrt{s}=7\TeV$
during the 2011 run of the Large Hadron Collider (LHC)~\cite{LHC:2008}.  The
ATLAS detector consists of an inner tracking detector surrounded by a thin
superconducting solenoid, electromagnetic (EM) and hadronic calorimeters, and a muon
spectrometer incorporating large superconducting toroid magnets. Each
subdetector is divided into barrel and end-cap components.

Only data taken with $pp$ collisions in stable beam conditions and with all
ATLAS subsystems operational are used, resulting in an integrated luminosity of
4.6~\ifb{}.  The data were collected using a combination of single-tau and
\ditau{} triggers, designed to select hadronic tau decays, and single-lepton
triggers. The \hadhad{} channel uses events passing either a \ditau{} trigger
with transverse energy (\et{}) thresholds of 20 and $29\GeV$, or a single-tau
trigger with an \et{} threshold of $125\GeV$.  The $\muhad{}$ and \emu{}
channels use events passing a single-muon trigger with a transverse momentum
(\pt{}) threshold of $18\GeV$, which is supplemented by accepting events that
pass a single-muon trigger with a \pt{} threshold of $40\GeV$ that operates
only in the barrel region but does not require a matching inner detector track.
The $\ehad$ channel uses events passing a single-electron trigger with \pt{}
thresholds in the range 20--$22\GeV$, depending on the data-taking period.
Events that pass the trigger are selected if the vertex with the largest sum of
the squared track momenta has at least four associated tracks, each with
\mbox{$\pt >0.5\GeV$}.

\begin{sloppypar} Monte Carlo (MC) simulation is used to estimate signal
efficiencies and some background contributions. MC samples of background
processes from \Wjets{} and \DYjets{} (enriched in high-mass \DY{}) events are
generated with {\tt ALPGEN~2.13}~\cite{Alpgen}, including up to five additional
partons.  Samples of $t\bar{t}$, $Wt$ and diboson ($WW$, $WZ$, and $ZZ$) events
are generated with {\tt MC@NLO~4.01}~\cite{MCNLO,Frixione:2008yi}.  For these
MC samples, the parton showering and hadronisation is performed by {\tt
HERWIG~6.520}~\cite{Corcella:2000bw} interfaced to {\tt
JIMMY~4.31}~\cite{Jimmy} for multiple parton interactions.  Samples of
$s$-channel and $t$-channel \singletop{} production are generated with {\tt
AcerMC~3.8}~\cite{AcerMC}, with the parton showering and hadronisation
performed by {\tt PYTHIA~6.425}~\cite{Pythia}.  Samples of $\ZPSSM$ signal
events are generated with {\tt PYTHIA~6.425}, for eleven mass hypotheses
ranging from 500 to $1750\GeV$ in steps of $125\GeV$.  In all samples photon
radiation is performed by {\tt PHOTOS}~\cite{PHOTOS}, and tau lepton decays are
generated with {\tt TAUOLA}~\cite{TAUOLA}.  The choice of parton distribution
functions (PDFs) depends on the generator: {\tt CTEQ6L1}~\cite{CTEQ6L1} is used
with {\tt ALPGEN}, {\tt CT10}~\cite{CT10} with {\tt MC@NLO} and {\tt
MRST2007~LO$^*$}~\cite{MRST2007} with {\tt PYTHIA} and {\tt AcerMC}.

The \DYonly{} cross section calculated at next-to-next-to-leading order (NNLO)
using {\tt PHOZPR}~\cite{PHOZPR} with {\tt MSTW2008} PDFs~\cite{Martin:2009} is
used to derive mass-dependent $K$-factors that are applied to the leading order
\DYjets{} and \ZP{} cross sections.  The \Wjets{} cross section is calculated
at NNLO using {\tt FEWZ 2.0}~\cite{WZprod,FEWZ2.0}.  The \ttbar{} cross section
is calculated at approximate NNLO~\cite{ttbarxsec1,ttbarxsec2,HATHOR}.  The
cross sections for single-top production are calculated at
next-to-next-to-leading-logarithm for the $s$-channel~\cite{schanxsec} and
approximate NNLO for $t$-channel and $Wt$ production modes~\cite{tchanxsec}. 

The detector response for each MC sample is simulated using a detailed {\tt
GEANT4}~\cite{geant4} model of the ATLAS detector and subdetector-specific
digitisation algorithms~\cite{AtlasSimulation}.  As the data are affected by
the detector response to multiple $pp$ interactions occurring in the same or in
neighbouring bunch crossings (referred to as pile-up), minimum-bias
interactions generated with {\tt PYTHIA~6.425} (with a specific LHC
tune~\cite{AMBT2}) are overlaid on the generated signal and background events.
The resulting events are re-weighted so that the distribution of the 
number of minimum-bias interactions per bunch crossing agrees with data. All
samples are  simulated with more than twice the effective luminosity of the
data, except $\Wjets$, where an equivalent of approximately $1.5\,\ifb$ is
simulated.  \end{sloppypar}


\section{Physics object reconstruction}
\label{sec:reconstruction_and_preselection}

\begin{sloppypar}
Muon candidates are reconstructed by combining an inner detector track
with a track from the muon spectrometer. They are required to have \mbox{$\pt{}
> 10\GeV$} and \mbox{$|\eta| < 2.5$}\footnote{
  ATLAS uses a right-handed coordinate system with its
  origin at the nominal interaction point (IP) in the centre of the detector and
  the $z$-axis along the beam pipe. The $x$-axis points from the IP to the centre
  of the LHC ring, and the $y$-axis points upward. Cylindrical coordinates
  $(r,\phi)$ are used in the transverse plane, $\phi$ being the azimuthal angle
  around the beam pipe. The pseudorapidity is defined in terms of the polar angle
  $\theta$ as $\eta=-\ln\tan(\theta/2)$. Separation in the $\eta$-$\phi$ plane 
  is defined as \mbox{$\Delta R = \sqrt{ (\Delta\eta)^2 + (\Delta\phi)^2 }$}.
}. Muon quality criteria 
are applied in order to achieve a precise measurement of the muon momentum and
reduce the misidentification rate~\cite{AtlasMuonEff2010}. These quality 
requirements correspond to a muon reconstruction and identification efficiency
of approximately 95\%. 
\end{sloppypar}   

\begin{sloppypar}
Electrons are reconstructed by matching clustered energy deposits in the
EM calorimeter to tracks reconstructed in the inner
detector~\cite{AtlasEgammaPerf2010}.  The electron candidates are required to
have \mbox{$\pt{} > 15\GeV$} and to be within the fiducial volume of the inner
detector, \mbox{$|\eta| < 2.47$}.  The transition region between the barrel and
end-cap EM calorimeters, with \mbox{$1.37 < |\eta| <1.52$}, is excluded.  The
candidates are required to pass quality criteria based on the expected
calorimeter shower shape and amount of radiation in the transition radiation
tracker.  These quality requirements correspond to an electron identification
(ID) efficiency of approximately 90\%. 
 Electrons and muons are considered
isolated   if they are away from large deposits of energy in the calorimeter, or
tracks with large \pt{} consistent with originating from the same vertex\footnote{
  Lepton isolation is defined using the sum of the \et{} deposited in
  calorimeter cells within $\Delta R < 0.2$ of the lepton, $\et^{0.2}$,
  and the scalar sum of the \pt{} of tracks with $\pt > 0.5\GeV$ consistent with the 
  same vertex as the lepton and within $\Delta R < 0.4$, $\pt^{0.4}$.
  Muons are considered isolated if they have $\et^{0.2}/\pt < 4\%$
  (and $\pt^{0.4}/\pt < 6\%$ in the $\taue\taumu$ channel).
  Isolated electrons must have $\pt^{0.4}/\pt < 5\%$ and $\et^{0.2}/\pt < 5\%$ if
  $\pt < 100 \GeV$ or $\et^{0.2} < 5 \GeV$ otherwise ($\et^{0.2}/\pt < 6\%$ and $\pt^{0.4}/\pt < 8\%$
  in the $\taue\taumu$ channel). 
}.
In the \ehad{} channel, isolated electrons are also required to pass a tighter
identification requirement corresponding to an efficiency of approximately
80\%.
\end{sloppypar}   

\begin{sloppypar} Jets are reconstructed using the \mbox{anti-$k_{t}$}
algorithm~\cite{anti-kt,Cacciari:2005hq} with a radius parameter value of 0.4.
The algorithm uses reconstructed, noise-suppressed clusters of calorimeter
cells~\cite{TopoClusters}.  Jets are calibrated to the hadronic energy scale
with correction factors based on simulation and validated using test-beam and
collision data~\cite{CERN-PH-EP-2011-191}.  All jets are required to have
\mbox{$\pt{} > 25\GeV$} and \mbox{$|\eta| < 4.5$}.  For jets within the inner
detector acceptance (\mbox{$|\eta|<2.4$}), the {\em jet vertex fraction} is
required to be at least 0.75; the jet vertex fraction is defined as the sum of
the \pt{} of tracks associated with the jet and consistent with originating
from the selected primary vertex, divided by the sum of the \pt{} of all tracks
associated with the jet.  This requirement reduces the number of jets that
originate from pile-up or are heavily contaminated by it.  Events are discarded
if a jet is associated with out-of-time activity or calorimeter
noise~\cite{ATLAS-CONF-2010-038}.  \end{sloppypar}   

\begin{sloppypar} Candidates for hadronic tau decays are defined as jets with
either one or three associated tracks reconstructed in the inner detector.  The
kinematic properties of the tau candidate are reconstructed from the {\em
visible} tau lepton decay products (all products excluding neutrinos).  The tau
charge is reconstructed from the sum of the charges of the associated tracks
and is required to be~$\pm1$. The charge misidentification probability is found
to be negligible.  Hadronic tau decays are identified with a multivariate
algorithm that employs boosted decision trees (BDTs) to discriminate against
quark- and gluon-initiated jets using shower shape and tracking
information~\cite{ATLAS-CONF-2011-077}.  Working points with a tau
identification efficiency of about 50\% ({\em medium}) for the $\muhad$ and
$\ehad$ channels and 60\% ({\em loose}) for the \hadhad{} channel are chosen,
leading to a rate of false identification for quark- and gluon-initiated jets
of a few percent~\cite{ATLAS-CONF-2011-152}.  Tau candidates are also required
to have \mbox{$\pt{} > 35\GeV$} and to be in the fiducial volume of the inner
detector, \mbox{$|\eta| < 2.47$} (the EM calorimeter transition region is
excluded).  In the $\lephad{}$ channels, tau candidates are required to have
only one track, which must not be in the range \mbox{$|\eta|<0.05$}, and to
pass a muon veto. The removed pseudorapidity region corresponds to a gap in the
transition radiation tracker that reduces the power of electron/pion
discrimination.  In the $\ehad{}$ channel, tau candidates are also required to
pass an electron veto using BDTs.  \end{sloppypar}   

\begin{sloppypar} Geometric overlap of objects with \mbox{$\Delta R < 0.2$} is
resolved by selecting only one of the overlapping objects in the following
order of priority: muons, electrons, tau candidates and jets.  The missing
transverse momentum (with magnitude \met{}) is calculated from the vector sum
of the transverse momenta of all high-\pt{} objects reconstructed in the event,
as well as a term for the remaining activity in the
calorimeter~\cite{AtlasMET2010}.  Clusters associated with electrons, hadronic
tau decays and jets are calibrated separately, with the remaining clusters
calibrated at the EM energy scale.  \end{sloppypar}

\section{Event selection}
\label{sec:event_selection}

\begin{sloppypar} 
Selected events in the \hadhad{} channel must contain at least two
oppositely-charged tau    candidates with \mbox{$\pt>50\GeV$} and no electrons
with \mbox{$\pt>15\GeV$} or muons with \mbox{$\pt>10\GeV$}.  If the event was
selected by the \ditau{} trigger, both tau candidates are required to be
geometrically matched to the objects that passed the trigger.  For events that
pass only the single-tau trigger there is no ambiguity, so trigger matching is
not required.  If multiple tau candidates are selected, the two highest-\pt{}
candidates are chosen.  The angle between the tau     candidates in the
transverse plane must be greater than 2.7\,radians.
\end{sloppypar}

\begin{sloppypar} 
Selected events in the $\lephad$ channels must contain exactly one isolated
muon with \mbox{$\pt>25\GeV$} or an isolated electron with \mbox{$\pt>30\GeV$};
no additional electrons with \mbox{$\pt{}>15\GeV$} or muons with
\mbox{$\pt{}>4\GeV$}; and exactly one tau     candidate with
\mbox{$\pt{}>35\GeV$}.  The angle between the lepton and tau candidate in the
transverse plane must be greater than 2.7\,radians, and the pair must have
opposite electric charge.  \end{sloppypar}

\begin{sloppypar}
For the $\ehad{}$ channel, the \mbox{$Z\to ee$} and multijet contributions 
are reduced to a negligible level by requiring $\met{}>30\GeV$.
The \mbox{$W$+jets} background is suppressed  by requiring 
the transverse mass, \mT{}, of the electron-\met{} system, defined as
\begin{equation}
m_{\rm T}=\sqrt{2p_{{\rm T}e}\metscalar(1-\cos\Delta\phi)} \, ,
\end{equation}
where $\Delta\phi$ is the angle between the lepton and \met{} in the 
transverse plane, to be less than $50\GeV$.
\end{sloppypar}

\begin{sloppypar}
Selected events in the \emu{} channel must contain exactly one isolated muon with
\mbox{$\pt>25\GeV$} and one isolated electron with \mbox{$\pt>35\GeV$} and 
opposite electric charge, no additional electrons with
\mbox{$\pt>15\GeV$} or muons with \mbox{$\pt>10\GeV$} and not more than one
jet. The jet requirement suppresses $\ttbar$ events, which typically have 
higher jet multiplicity than the signal.  
The two leptons are required to be
back-to-back in the transverse plane using the criterion $\pzetavis < 10\GeV$, with
\begin{equation}
\pzetavis = \vec{p}_{{\rm T}e}\cdot\hat{\zeta} + \vec{p}_{{\rm T}\mu}\cdot\hat{\zeta} \, ,
\end{equation}
where $\hat{\zeta}$ is a unit vector along the bisector of the $e$ and $\mu$ momenta.
This selection provides good suppression of the \diboson{} and \ttbar{} backgrounds. 
For \Zprime{} events, the \met{} tends to point away from the highest-\pt{} lepton, 
so the angle between the highest-\pt{} lepton and \met{} in
the transverse plane is required to be greater than 2.6\,radians. 
\end{sloppypar}

The search in all channels is performed by counting events in signal regions
with {\em total transverse mass} above thresholds optimised separately for each
signal mass hypothesis in each channel to give the best expected exclusion
limits (see Table~\ref{tab:MT_thresholds}).  
The total transverse mass, \MT, is
defined as the mass of the visible decay products of both tau leptons and
\met{}, neglecting longitudinal momentum components and the tau lepton mass, 
\begin{equation}
\hspace{-6mm}\MT = \sqrt{2\ptone\pttwo\Conetwo + 2\metscalar\ptone\Conemiss + 2\metscalar\pttwo\Ctwomiss}\, , 
\end{equation}
where \ptone{} and \pttwo{} are the transverse momenta of the visible products 
of the two tau decays; \Conetwo{} is defined as $1-\cos\Delta\phi$, where $\Delta\phi$ is
the angle in the transverse plane between the visible products of the two tau decays; and
\Conemiss{} and \Ctwomiss{} are defined analogously for the angles in the transverse plane 
between \met{} and the visible products of the first and second tau decay,
respectively.
Figs.~\ref{fig:finalmass}(a)--(d) show the \MT{} distribution after event
selection in each channel.

\begin{table}[htbp]
    \small
    \centering
    \begin{tabular}{lrrrrrrr}
        \hline\hline
        $m_{\Zprime{}}$ & 500 & 625 & 750 & 875 & 1000 & 1125 & $\geq$1250 \\ 
        \hline                                                                                    
        \tauh\tauh      & 350 & 400 & 500 & 500 &  650 &  650 & 700        \\
        $\taumu\tauh$   & 400 & 400 & 500 & 500 &  600 &  600 & 600        \\
        $\taue\tauh$    & 400 & 400 & 400 & 500 &  500 &  500 & 500        \\
        $\taue\taumu$   & 300 & 350 & 350 & 350 &  500 &  500 & 500        \\
        \hline\hline
    \end{tabular}
    \caption{
        Thresholds on $\MT{}$ used for each signal mass point in each channel. 
        All values are given in GeV. 
    }
    \label{tab:MT_thresholds}
\end{table}

\begin{figure*}[t!]
    \centering
    \includegraphics[width=0.40\textwidth]{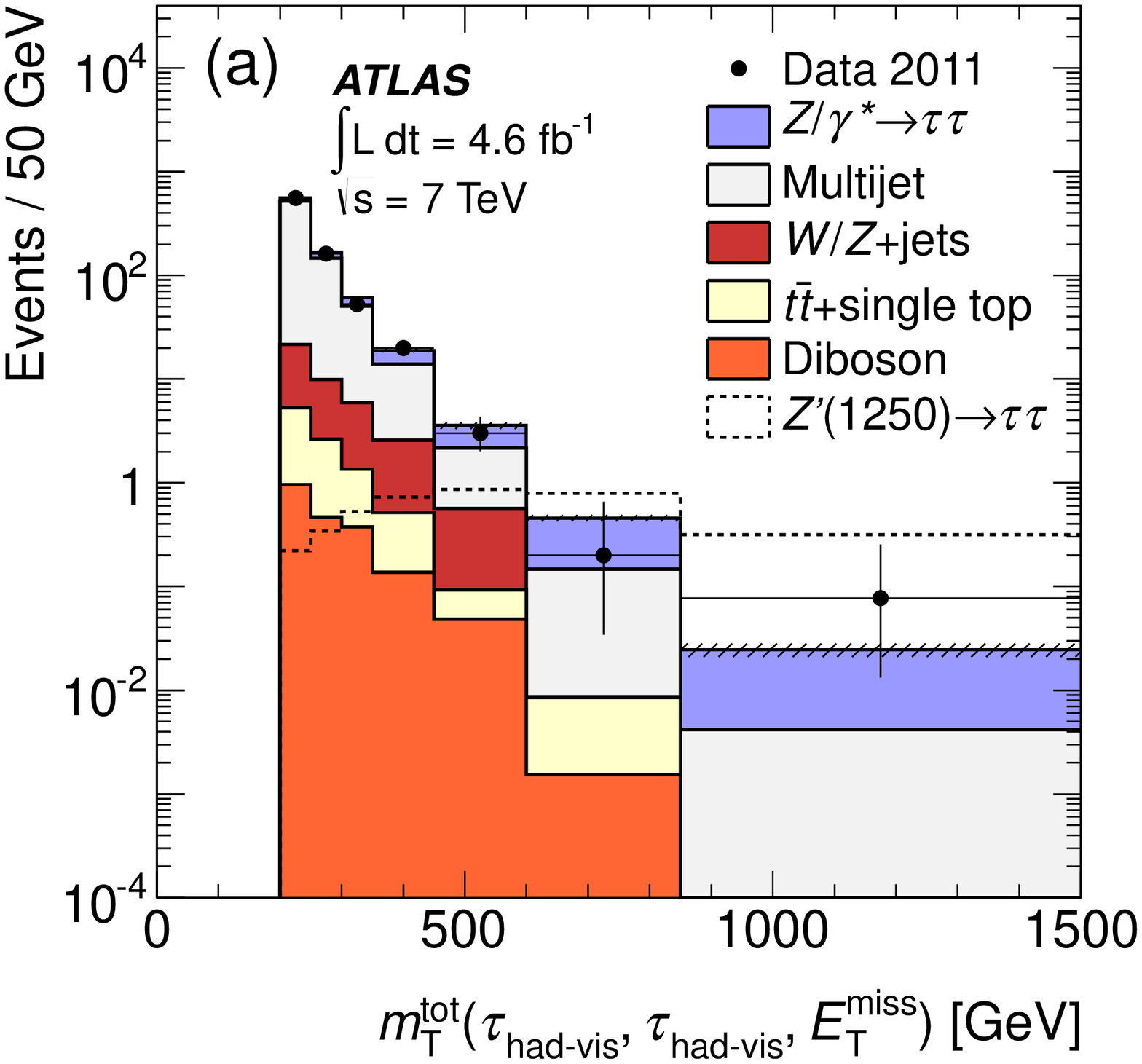}
    \includegraphics[width=0.40\textwidth]{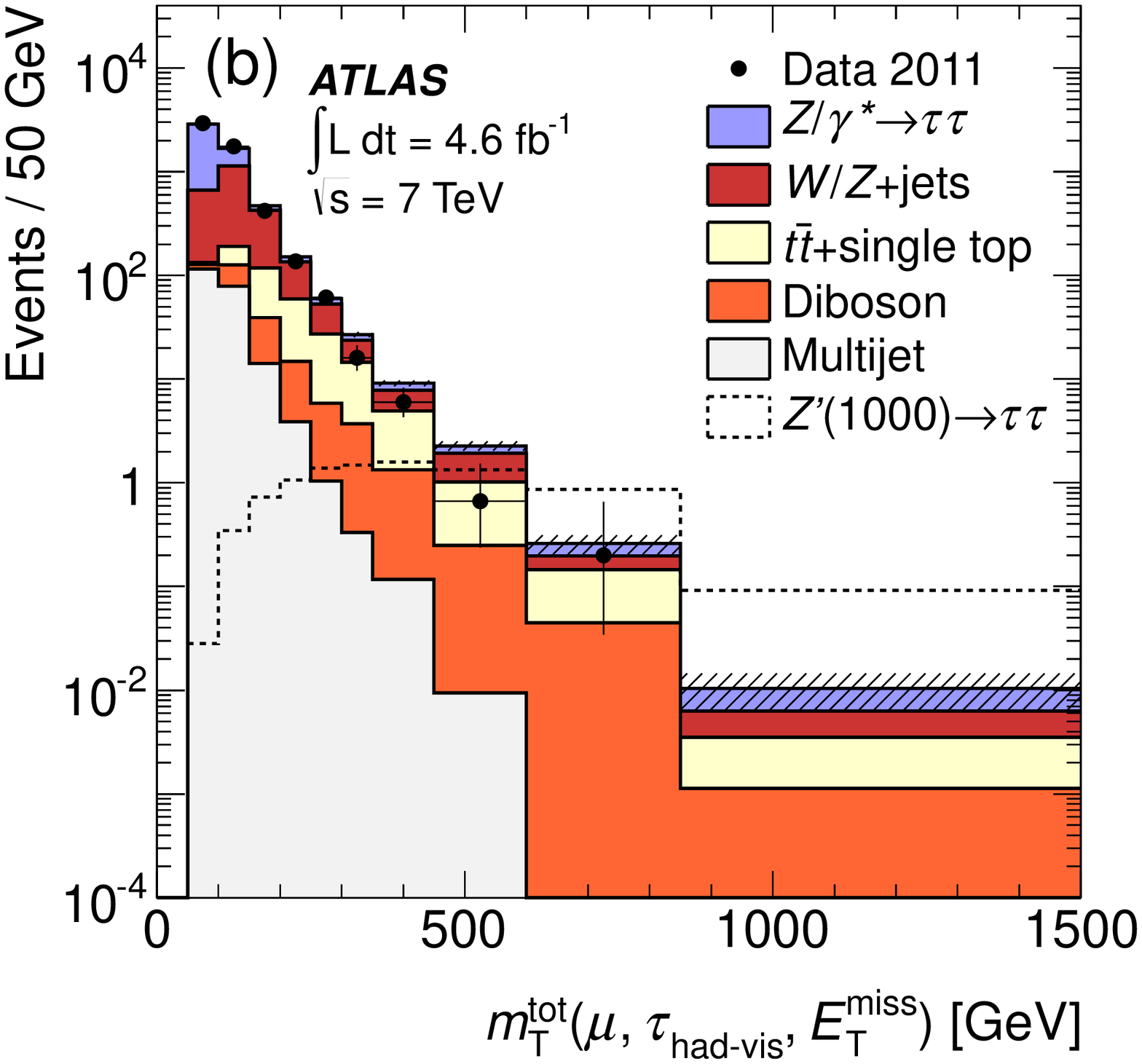}
    \includegraphics[width=0.40\textwidth]{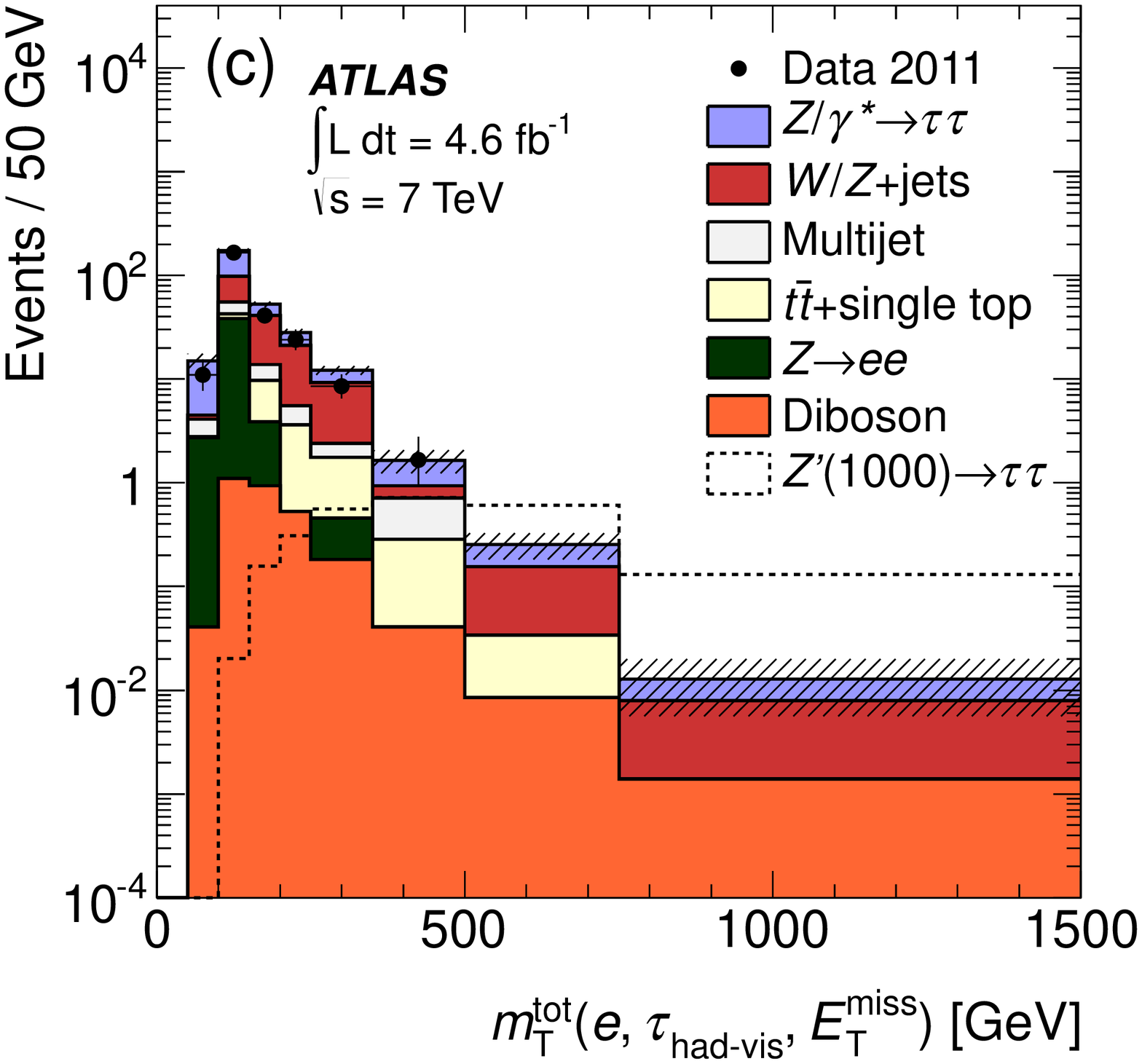}
    \includegraphics[width=0.40\textwidth]{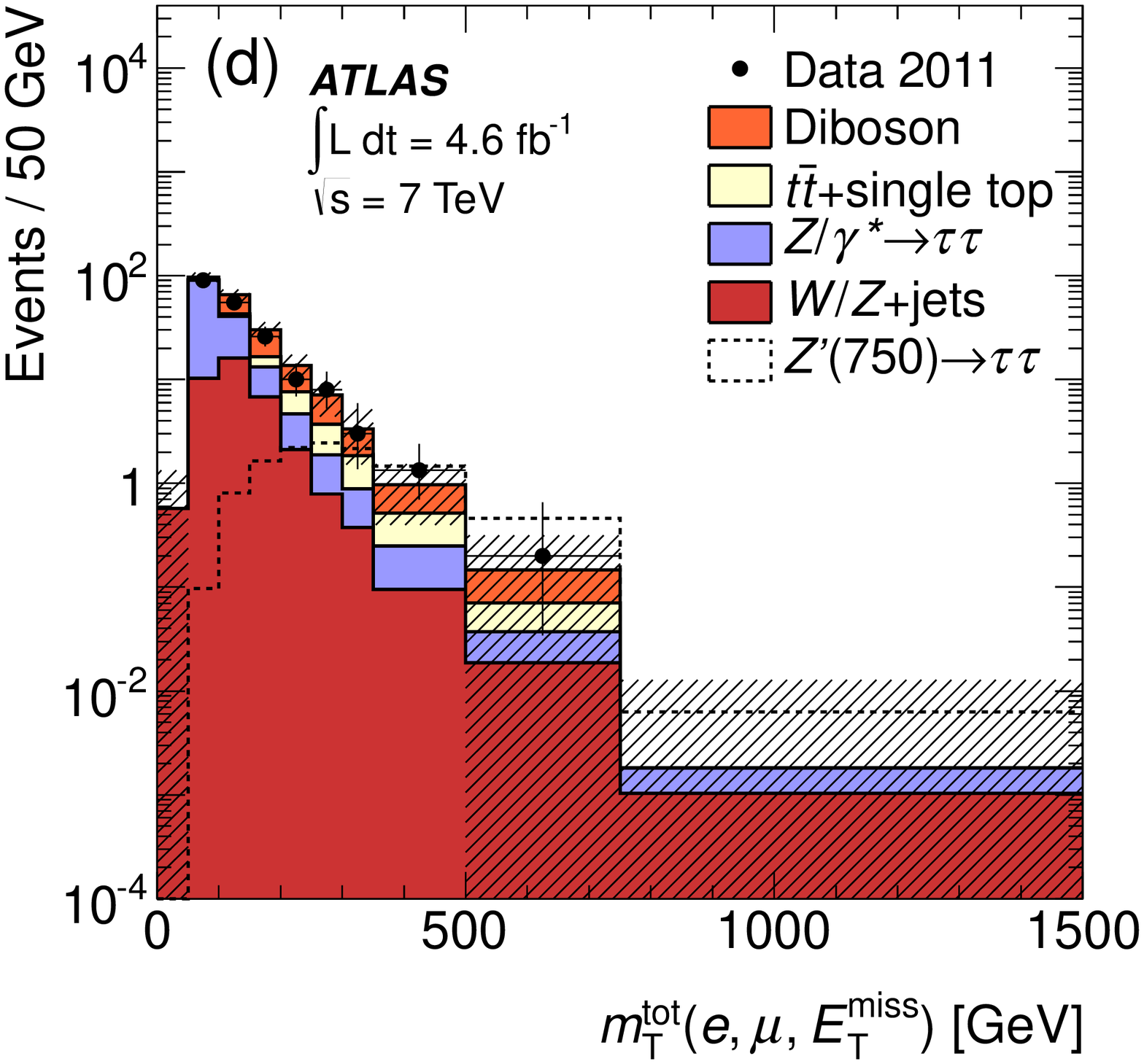}
    \caption{
        The \MT{} distribution after event selection without the \MT{} requirement for
        each channel: (a) \hadhad{}, (b) $\muhad{}$, (c) $\ehad{}$ and (d) \emu{}. The
        estimated contributions from SM processes are stacked and appear in the same
        order as in the legend. The contribution from $Z\to ee$ events in which an
        electron is misidentified as a tau candidate is shown separately in the \ehad{}
        channel. A \ZprimeSSM{} signal and the events observed in data are overlaid.
        The signal mass point closest to the \ZprimeSSM{} exclusion limit in each
        channel is chosen and is indicated in parentheses in the legend in units
        of$\GeV$.  The uncertainty on the total estimated background (hatched) includes
        only the statistical uncertainty from the simulated samples. The visible decay
        products of hadronically decaying taus are denoted by \tauhvis{}. 
    }
    \label{fig:finalmass}
\end{figure*}

\section{Background estimation}
\label{sec:bkg_est}

\begin{sloppypar}
The dominant background processes in the \hadhad{} channel are \multijet{}
production and \DY{}.  Minor contributions come from \mbox{\Wtaunujets{}},
\mbox{\Zlljets{}} ($\ell=e$ or $\mu$), \mbox{\Wlnujets{}}, \ttbar{},
\singletop{} and \diboson{} production. The shape of the \multijet{} mass distribution is
estimated from data that pass the full event selection but have two tau
candidates of the same electric charge. The contribution is normalised to
events that pass the full event selection but have low \MT{}.  All other
background contributions are estimated from simulation.  \end{sloppypar}

\begin{sloppypar}
The main background contributions in the \lephad{} channels come from
\mbox{\DY{}}, \mbox{\Wjets{}}, \ttbar{} and \diboson{} production, with minor
contributions from \mbox{\Zlljets{}}, \multijet{} and \singletop{} events.  The
contributions involving fake hadronic tau decays from multijet and
\mbox{\Wjets{}} events are modelled with data-driven techniques involving {\em fake
factors}, which parameterise the rate for lepton candidates in jets to pass lepton
isolation or jets to pass tau identification, respectively.  The remaining
background is estimated using simulation.
\end{sloppypar}

The dominant background processes in the \emu{} channel are \ttbar{},
\mbox{\DY{}} and \diboson{} production. Contributions from processes such as
\mbox{\Zmumujets{}}, \mbox{\Wjets{}} and \mbox{\Wgammajets{}}, where a jet or
photon is misidentified as an electron, are very small in the signal region.
Multijet events are suppressed by tight lepton isolation criteria. Since
background processes involving fake leptons make only minor contributions, all
background contributions in the \emu{} channel are estimated using simulation.
The MC estimates of the dominant background contributions are checked using
high-purity control regions in data.

The following subsections describe the data-driven background estimates in more detail.

\subsection{Multijet background in the \hadhad{} channel}
\label{sec:had_had_background_estimation}
\begin{sloppypar}
The shape of the \MT{} distribution for the \multijet{} background is estimated
using events that pass the standard event selection, but have two selected
\tauh{} candidates with the same electric charge and with \mbox{$\MT>200\gev$}
to avoid the low \MT{} region which is affected by the tau \pt{} threshold.
For a low-mass signal with $m_{Z'}\leq625\GeV$, a
lower bound of $160\GeV$ is used, as discussed below.  This control region has only 2\%
contamination from other background processes and negligible signal contamination.  The
\MT{} distribution is modelled by performing an unbinned maximum likelihood fit
to the data in the control region using the following function:
\begin{equation}
f(\MT{}|p_0,p_1,p_2) = p_0 \cdot \left(\MT{}\right)^{p_1+p_2\log{\left(\MT{}\right)}},
\end{equation}
where $p_0$, $p_1$ and $p_2$ are free parameters. 
The integral of the fitted function in the high-mass tail matches the number of
observed events well for any choice of the \MT{} threshold, and the function
models the high-mass tail well in a simulated dijet sample enriched in
high-mass events.  The statistical uncertainty is estimated using
pseudo-experiments and increases monotonically from 12\% to 83\% with
increasing \MT{} threshold. The systematic uncertainty due to the choice of the
fitting function is evaluated using alternative fitting functions and ranges
from 1\% to 7\%.  The \multijet{} model is normalised to data that pass all
analysis requirements but have \MT{} in the range \mbox{$200$--$250\GeV$}.  For
the low-mass points with \mbox{$m_{Z'}\leq625\GeV$}, the low-\MT{} side-band is
lowered to \mbox{$160$--$200\GeV$} to keep signal contamination negligible.
Both side-bands have a maximum contamination of 5\% from other background processes,
which is subtracted, and negligible contamination from signal. The statistical
uncertainty from the normalisation ranges from 2\% to 5\%.  Systematic
uncertainties affecting the normalisation of the background processes are
propagated when performing the subtraction but have a negligible effect.
\end{sloppypar} 

\subsection{Multijet background in the $\lephad{}$ channels}
\label{sec:mu_had_bkg_qcd}
\label{sec:e_had_bkg_qcd}

The background from multijet events is negligible at high \MT{}, but it is
important to estimate its contribution to model the inclusive mass distribution.
Multijet events are exceptional among the background processes because
the muons and
electrons produced in heavy-flavour decays or the light-flavour hadrons falsely
identified as electrons, are typically not isolated in the calorimeter
but produced in jets.
To estimate the multijet background, events in the data that
fail lepton isolation are weighted event-by-event, with fake factors for lepton
isolation measured from data in a multijet-rich control region (multijet--CR).  The
multijet--CR is defined by requiring exactly one
selected lepton, as in Section~\ref{sec:event_selection}, but without the isolation requirement; 
at least one tau candidate that fails the BDT ID; no tau candidates that pass the BDT ID; 
$\met{} < 15 \GeV$ for the \muhad{}
channel, $\met{} < 30 \GeV$ for the $\ehad{}$ channel; and the transverse mass
formed by the lepton and \met{}, $\mt(\ell, \met)$, to be less
than $30\GeV$.  For the \muhad{} channel, where the multijet contribution
is dominated by $b$-quark-initiated jets, the muon is additionally required
to have a transverse impact parameter of $|d_0(\mu)|>0.08$\,mm
with respect to the primary vertex, 
which increases the purity of the multijet control region.  The
leptons in the multijet control region are divided into those that pass
(isolated) and a subset that fail (anti-isolated) the isolation requirements.
In the \muhad{} channel the anti-isolated sample includes all muons that 
fail isolation, while in the \ehad{} channel, the anti-isolation requirement 
is tightened to reduce contamination from real isolated electrons.
Isolation fake factors, $f_{\mathrm{iso}}$, are defined as the number of
isolated leptons in the data, \Niso{}, divided by the number of anti-isolated
leptons, \Nantiiso{}, binned in $\pt$ and $\eta$:
\begin{equation}
    f_{\mathrm{iso}}(\pt,\eta) \equiv \left.\frac{\Niso(\pt, \eta)}{\Nantiiso(\pt, \eta)}\right|_\mathrm{multijet-CR} \, .
\end{equation}
Contamination from real isolated leptons is estimated using simulation and
subtracted from $\Niso$ ($\sim$3\% for \muhad{} and $\sim$25\% for \ehad{}).
The number of multijet events passing lepton isolation, $N_\mathrm{multijet}$,
is predicted by weighting the events with anti-isolated leptons by their fake
factor:
\begin{eqnarray}
\hspace{-5mm}
\nonumber N_\mathrm{multijet}(\pt,\eta, x)  =  f_{\mathrm{iso}}(\pt,\eta) \qquad \qquad \qquad \qquad && \\
  \times \Big( \Nantiiso_\mathrm{data}(\pt, \eta, x) - \Nantiiso_\mathrm{MC}(\pt, \eta, x) \Big)\, . &&
\end{eqnarray}
The shape of the multijet background 
in a given kinematic variable, $x$, 
is modelled from the events in the data with anti-isolated leptons,
$\Nantiiso_\mathrm{data}$, corrected by subtracting the expected contamination
from other background processes predicted with MC simulation, $\Nantiiso_\mathrm{MC}$.

This method assumes that the ratio of the number of isolated leptons to the
number of anti-isolated leptons in multijet events is not strongly correlated
with the requirements used to enrich the multijet control sample.  This
assumption has been verified by varying the $\met$ and $d_0$ selection criteria used to
define the multijet control region. A conservative 100\% systematic
uncertainty on the isolation fake factor is assumed, but this has negligible
effect on the sensitivity because the expected multijet background is less than
a percent of the total background in both the \muhad{} and \ehad{} channels.

\subsection{$W$+jets background in the $\lephad{}$ channels}
\label{sec:mu_had_bkg_wjet}
\label{sec:e_had_bkg_wjet}

\begin{sloppypar}
The \Wjets{} background is estimated using a technique similar to the 
\multijet{} estimate, where tau candidates that fail the BDT ID 
are weighted event-by-event with fake factors for jets
to pass the BDT ID in \Wjets{} events.  A high purity \Wjets{} control
region (W--CR) is defined by selecting events that have
exactly one isolated lepton, as in Section~\ref{sec:event_selection};
at least one tau candidate that is not required to pass the BDT ID;
and $\mt(\ell, \met)$ between 70 and $200\GeV$.
For the \ehad{} channel, the tau candidate is additionally
required to pass the electron veto.
Tau ID fake factors, $f_{\tau}$, are defined as the number of 
tau candidates that pass the BDT ID,
\mbox{$N^{\mathrm{pass\ }\tau\mathrm{-ID}}$},
divided by the number that fail,
\mbox{$N^{\mathrm{fail\ }\tau\mathrm{-ID}}$},
binned in $\pt$ and $\eta$:
\begin{equation}
    f_{\tau}(\pt,\eta) \equiv \left.\frac{N^{\mathrm{pass\ }\tau\mathrm{-ID}}(\pt, \eta)}{N^{\mathrm{fail\ }\tau\mathrm{-ID}}(\pt, \eta)}\right|_\mathrm{W-CR}   \, .
\end{equation}
The number of \Wjets{} events passing the BDT ID, $N_{W+\mathrm{jets}}$, is predicted by
weighting the events that fail the BDT ID by their fake factor:
\begin{eqnarray}
\hspace{-7mm}
\nonumber N_{W+\mathrm{jets}}(\pt,\eta, x) = f_{\tau}(\pt,\eta) \times \Big( N_\mathrm{data}^{\mathrm{fail\ }\tau\mathrm{-ID}}(\pt, \eta, x)  && \\
    - \: N_\mathrm{multijet}^{\mathrm{fail\ }\tau\mathrm{-ID}}(\pt, \eta, x)
    - N_\mathrm{MC}^{\mathrm{fail\ }\tau\mathrm{-ID}}(\pt, \eta, x) \Big)\, .  &&
\end{eqnarray}
The shape of the \mbox{$W$+jets} background is modelled using events in the
data that failed the BDT ID, $ N_\mathrm{data}^{\mathrm{fail\
}\tau\mathrm{-ID}}$, with the multijet contamination,
$N_\mathrm{multijet}^{\mathrm{fail\ }\tau\mathrm{-ID}}$ (estimated from data),
and other contamination, $N_\mathrm{MC}^{\mathrm{fail\ }\tau\mathrm{-ID}}$
(estimated from simulation), subtracted.
\end{sloppypar}

A 30\% systematic uncertainty on the fake factors is assigned 
by comparing the fake factors to those measured in a data sample enriched in \mbox{$Z$+jets}
instead of \Wjets{}, which provides a sample of jets with a similar
quark/gluon fraction~\cite{Gallicchio:2011-JHEP}.
This background estimation method relies on the assumption that the tau
identification fake factors for \Wjets{} events are not strongly correlated with
the selection used to define the \Wjets{} control region. This assumption has been
verified by varying the \mT{} selection criterion used to define the \Wjets{} control region,
resulting in a few percent variation, which is well within the systematic uncertainty.

\section{Systematic uncertainties}
\label{sec:common_systematics}

Systematic effects on the contributions of signal and background processes
estimated from simulation are discussed in this section.  These include
theoretical uncertainties on the cross sections of simulated processes and
experimental uncertainties on the trigger, reconstruction and identification
efficiencies; on the energy and momentum scales and resolutions; and on the
measurement of the integrated luminosity.  For each source of uncertainty, the
correlations across analysis channels, as well as the correlations between
signal and background, are taken into account.  Uncertainties on the background
contributions estimated from data have been  discussed in their respective
sections.

\begin{sloppypar}
The overall uncertainty on the \Zprime{} signal and the
\hbox{\DY{}} background due to PDFs, $\alpha_{S}$ and scale variations is
estimated to be 12\% at 1.5~TeV, dominated by the PDF
uncertainty~\cite{ATLAS-CONF-2012-007}.  The uncertainty is evaluated using PDF
error sets, and the spread of the variations covers the difference between the
central values obtained with the {\tt CTEQ} and {\tt MSTW} PDF sets.
Additionally, for \DY{}, a systematic uncertainty of 10\% is attributed to
electroweak corrections~\cite{AtlasZprime2011}. This uncertainty is not
considered for the signal as it is strongly model-dependent. An uncertainty of
4--5\% is assumed for the inclusive cross section of the single gauge boson and
\diboson{} production mechanisms and a relative uncertainty of 24\% is added in
quadrature per additional jet, due to the irreducible Berends-scaling
uncertainty~\cite{Berends1989,Berends1991}.  For \ttbar{} and \singletop{}
production, the QCD scale uncertainties are in the range of
3--6\%~\cite{tchanxsec,PhysRevD78034003,Beneke}. The uncertainties
related to the proton PDFs, including those arising from the choice of PDF set,
amount to 8\% for the predominantly gluon-initiated
processes such as \ttbar\ and 4\% for the predominantly quark-initiated
processes at low mass, such as on-shell single gauge boson and \diboson{}
production~\cite{CT10,Martin:2009,Botje:2011sn,2011arXiv1101.0536A,Ball:2011mu}.
\end{sloppypar}

The uncertainty on the integrated luminosity is
3.9\%~\cite{AtlasLumi:2011,Aad:2011dr}.
The efficiencies of the electron, muon and hadronic tau
triggers are measured in data and are used to correct the simulation. The
associated systematic uncertainties are typically \mbox{1--2\%} for electrons
and muons, \mbox{2.5\%} for the \ditau{} trigger and \mbox{5\%} for the high-\pt{}
single-tau trigger.  Differences between data and simulation in
the reconstruction and identification efficiencies of electrons, muons, and
hadronic tau decays are taken into account, as well as the differences in the
energy and momentum scales and resolutions.  The associated uncertainties for muons and
electrons are typically $<$1\%.

The systematic uncertainties on the identification efficiency of hadronic tau
decays are estimated at low \pt{} from data samples enriched in \Wtaunu{} and
\Ztautau{} events.  At high \pt{}, there are no abundant sources of real
hadronic tau decays to make an efficiency measurement. Rather, the fraction of
jets that pass the tau identification is studied in high-\pt{} dijet events as
a function of the jet \pt{}, which indicates that there is no degradation in
the modelling of the detector response as a function of the \pt{} of tau
candidates.  From these studies, an efficiency uncertainty of up to 8\% is
assigned to high-\pt{} tau candidates.  The uncertainty on the jet-to-tau
misidentification rate is 50\%, determined from data-MC comparisons in $W$+jet
events.  The uncertainty on the electron-to-tau misidentification rate is
50--100\%, depending on the pseudorapidity of the tau candidate, based on
measurements made using a $Z\to ee$ sample selected from data
\cite{ATLAS-CONF-2011-152}.  The energy scale uncertainty on taus and jets is
evaluated based on the single-hadron response in the
calorimeters~\cite{CERN-PH-EP-2011-191,ATLAS-CONF-2012-054}. In addition, the
tau energy scale is validated in data samples enriched in \Ztautau{} events.
The systematic uncertainties related to the jet and tau energy scale and
resolution are functions of $\eta$ and \pt{}, and are generally near 3\%. These
uncertainties are treated as fully correlated.  Energy scale and resolution
uncertainties on all objects are propagated to the \met{} calculation.  The
uncertainty on the \met{} due to clusters that do not belong to any
reconstructed object is measured to be negligible in all channels. 

Table~\ref{tab:finalsys} summarises the uncertainties on the estimated signal
and total background contributions in each channel. For the background, the
contribution from each uncertainty depends on the fraction of the background
estimated with simulation.  The dominant uncertainties on the background come
from the multijet shape estimation and the tau energy scale uncertainty for
\DY{} events in the \hadhad{} channel, from the limited sample size and the
fake factor estimate of the $W$+jets background in the \lephad{} channels and
from the statistical uncertainty of the MC samples in the $\emu$ channel.  The
dominant uncertainty on the signal for the $\hadhad{}$ and \lephad{} channels
comes from the tau identification efficiency and for the $\emu$~channel, from
the statistical uncertainty on the MC samples.

\begin{table}[htbp]
    \renewcommand{\tabcolsep}{1mm}
    \small
    \centering
    \begin{tabular}{lrrrrrrrrrr}
        \hline\hline
Uncertainty [\%]                    & \hspace{1mm}     & \multicolumn{4}{c}{Signal} &\hspace{3mm}                    & \multicolumn{4}{c}{Background}   \\
                                    &      & hh &$\mu$h &$e$h & $e\mu$ &  & hh &$\mu$h &$e$h & $e\mu$ \\ 
        \hline                                                                                     
        Stat. uncertainty           &      &  1 &  2 &  2 &  3 &  &              5 & 20 & 23 &  7 \\
        Eff. and fake rate          &      & 16 & 10 &  8 &  1 &  &             12 & 16 &  4 &  3 \\
        Energy scale and res.       &      &  5 &  7 &  6 &  2 &  & $^{+22}_{-11}$ &  3 &  8 &  5 \\
        Theory cross section        &      &  8 &  6 &  6 &  5 &  &              9 &  4 &  4 &  5 \\
        Luminosity                  &      &  4 &  4 &  4 &  4 &  &              2 &  2 &  2 &  4 \\
        Data-driven methods         &      & -- & -- & -- & -- &  & $^{+21}_{-11}$ &  6 & 16 & -- \\
        \hline\hline
    \end{tabular}
    \caption{
        Uncertainties on the estimated signal and total background contributions in
        percent for each channel.  
        The following signal masses, chosen to be close to the region where the limits are set,
        are used: $1250\GeV$ for $\hadhad{}$ (hh); $1000\GeV$ for $\muhad{}$ ($\mu$h) and $\ehad$ ($e$h);
        and $750\GeV$ for $\emu$ ($e\mu$).
        A dash denotes that the uncertainty is not applicable.  The statistical
        uncertainty corresponds to the uncertainty due to limited sample size in the MC
        and control regions.
    }
    \label{tab:finalsys}
\end{table}

\section{Results and discussion}
\label{sec:stats_results}
The numbers of observed and expected events including their total uncertainties,
after the full selection in all channels, are summarised in
Table~\ref{tab:eventsummary}. In all cases, the number of observed events is
consistent with the expected Standard Model background.  Therefore, upper limits
are set on the production of a high-mass resonance decaying to \tptm{} pairs.

\begin{table*}[htb!]
    \centering
    \begin{tabular}{lrrrr}
    \hline
    \hline
                                  & \hadhad{}      & \muhad{}       & \ehad{}        & \emu{} \\
    $m_{\Zprime}$     [GeV]       & 1250           & 1000           & 1000           & 750    \\
    $\MT{}$ threshold [GeV]       & 700            & 600            & 500            & 350    \\
    \hline                                                                                       
    \DY{}                         & 0.73$\pm$0.23  & 0.36$\pm$0.06  & 0.57$\pm$0.11  & 0.55$\pm$0.07\\
    \Wjets{}                      & $<0.03$        & 0.28$\pm$0.22  & 0.8 $\pm$0.4   & 0.33$\pm$0.10\\
    \Zlljets{}                    & $<0.01$        & $<0.1$         & $<0.01$        & 0.06$\pm$0.02\\
    \ttbar{}                      & $<0.02$        & 0.33$\pm$0.15  & 0.13$\pm$0.09  & 0.97$\pm$0.22\\
    \Diboson                      & $<0.01$        & 0.23$\pm$0.07  & 0.06$\pm$0.03  & 1.69$\pm$0.24\\
    Single top                    & $<0.01$        & 0.19$\pm$0.18  & $<0.1$         & $<0.1$ \\
    Multijet                      &  0.24$\pm$0.15 & $<0.01$        & $<0.1$         & $<0.01$ \\
    \hline                                                                             
    Total expected background     &  0.97$\pm$0.27 & 1.4 $\pm$0.4   & 1.6 $\pm$0.5   & 3.6 $\pm$0.4 \\
    Events observed               &              2 & 1              & 0              & 5\\
    \hline                                                                           
    Expected signal events        &  6.3 $\pm$1.1  & 5.5 $\pm$0.7   & 5.0 $\pm$0.5   & 6.72$\pm$0.26 \\
    Signal efficiency (\%)        &  4.3           & 1.1            & 1.0            & 0.4  \\
    \hline
    \hline
    \end{tabular}
    \caption{
        Number of expected and observed events after event selection for each
        analysis channel.  The expected contribution from the signal and background 
        in each channel is calculated for the signal mass point closest to the 
        \ZprimeSSM{} exclusion limit.
        The total uncertainties on each estimated contribution are shown.
        The signal efficiency denotes the expected number of signal events divided by the product
        of the production cross section, the ditau branching fraction and the integrated luminosity,
        $\sigma(pp\to Z'_{\rm SSM})\times {\rm BR}(\ZprimeSSM\to\tau\tau) \times \int L dt$. 
    }
    \label{tab:eventsummary}
\end{table*}

The statistical combination of the channels employs a likelihood function
constructed as the product of Poisson probability terms describing the total
number of events observed in each channel. The Poisson probability in each
channel is evaluated for the observed number of data events given the signal
plus background expectation. Systematic uncertainties on the expected number of
events are incorporated into the likelihood via Gaussian-distributed nuisance
parameters. Correlations across channels are taken into account.  A signal
strength parameter multiplies the expected signal in each channel, for which a
positive uniform prior probability distribution is assumed. Theoretical 
uncertainties on the signal cross section are not included in the calculation
of the experimental limit as they are model-dependent.

Bayesian 95\% credibility upper limits are set on the cross section 
times branching fraction for a high-mass resonance
decaying into a $\tau^{+}\tau^{-}$ pair as a function of the resonance mass,
using the Bayesian Analysis Toolkit~\cite{BAT}.  Figs.~\ref{fig:results}(a)
and \ref{fig:results}(b) show the limits for the individual channels and for
the combination, respectively.  The resulting 95\% credibility lower limit on the mass of a \ZprimeSSM{} 
decaying to \tptm{} pairs is $1.40\TeV$, with an expected limit of $1.42\TeV$. 
The observed and expected limits in the individual channels are, respectively:
1.26 and 1.35~\TeV{} (\hadhad{}); 1.07 and 1.06~\TeV{} ($\muhad{}$); 1.10
and 1.03~\TeV{} ($\ehad{}$); and 0.72 and 0.82~\TeV{} ($\emu$).

The impact of the choice of the prior on the signal strength parameter has been
evaluated by also considering the reference prior~\cite{RefPriors}.  Use
of the reference prior improves the mass limits by approximately $50\GeV$.  The
impact of the vector and axial coupling strengths of the $\Zprime{}$ has been
investigated, as these can alter the fraction of the tau momentum carried by the
visible decay products. For purely $V-A$ couplings, the limit on the cross section
times \tptm{} branching fraction is improved by $\sim$10\% over the mass range.
For purely $V+A$ couplings, there is a mass-dependent degradation, from $\sim$15\% at
high mass to $\sim$40\% at low mass.  All variations lie within the $1\sigma$
band of the expected exclusion limit.

\begin{figure*}[htb!]
    \centering
    \includegraphics[width=0.48\textwidth]{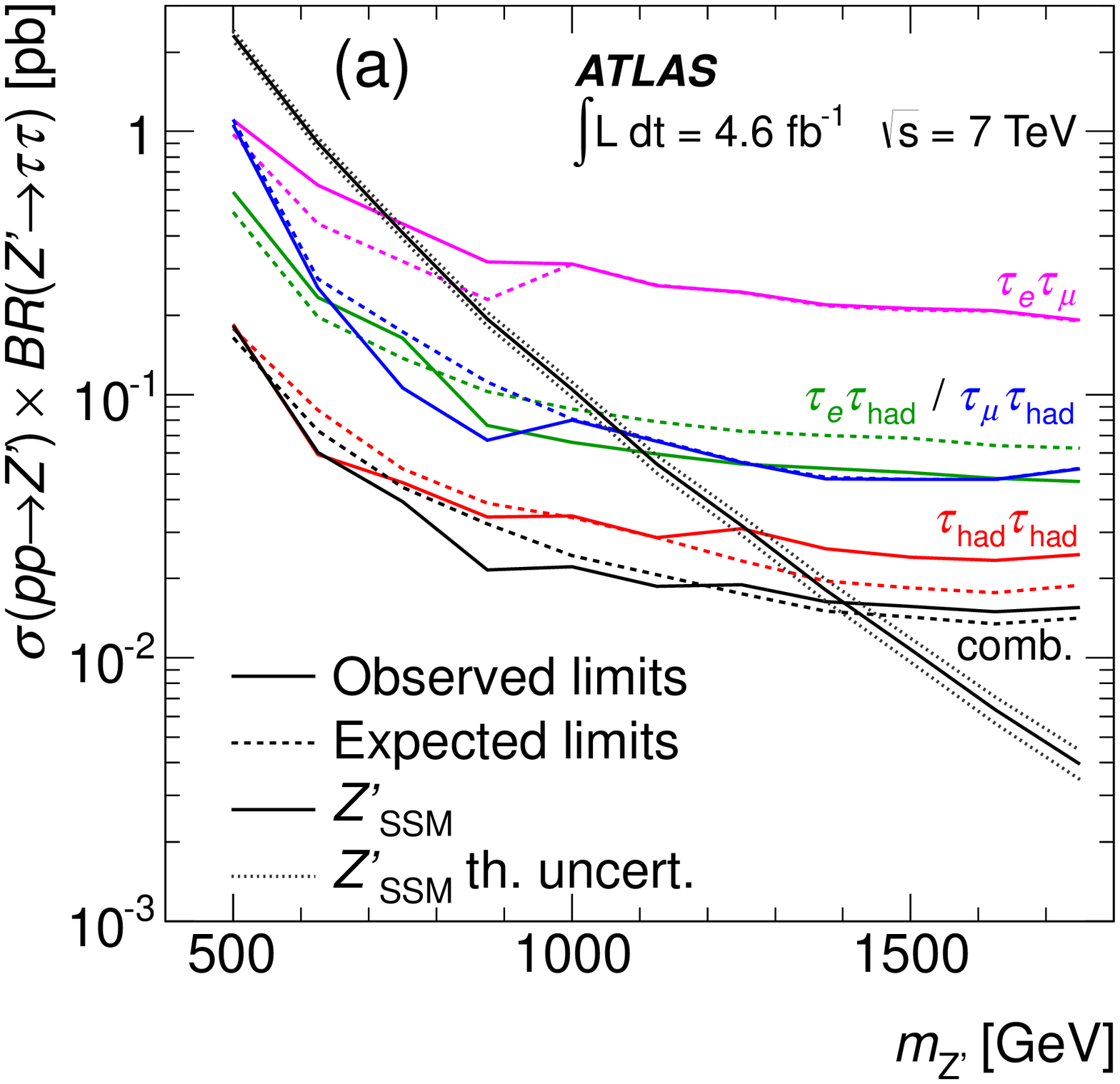}
    \includegraphics[width=0.48\textwidth]{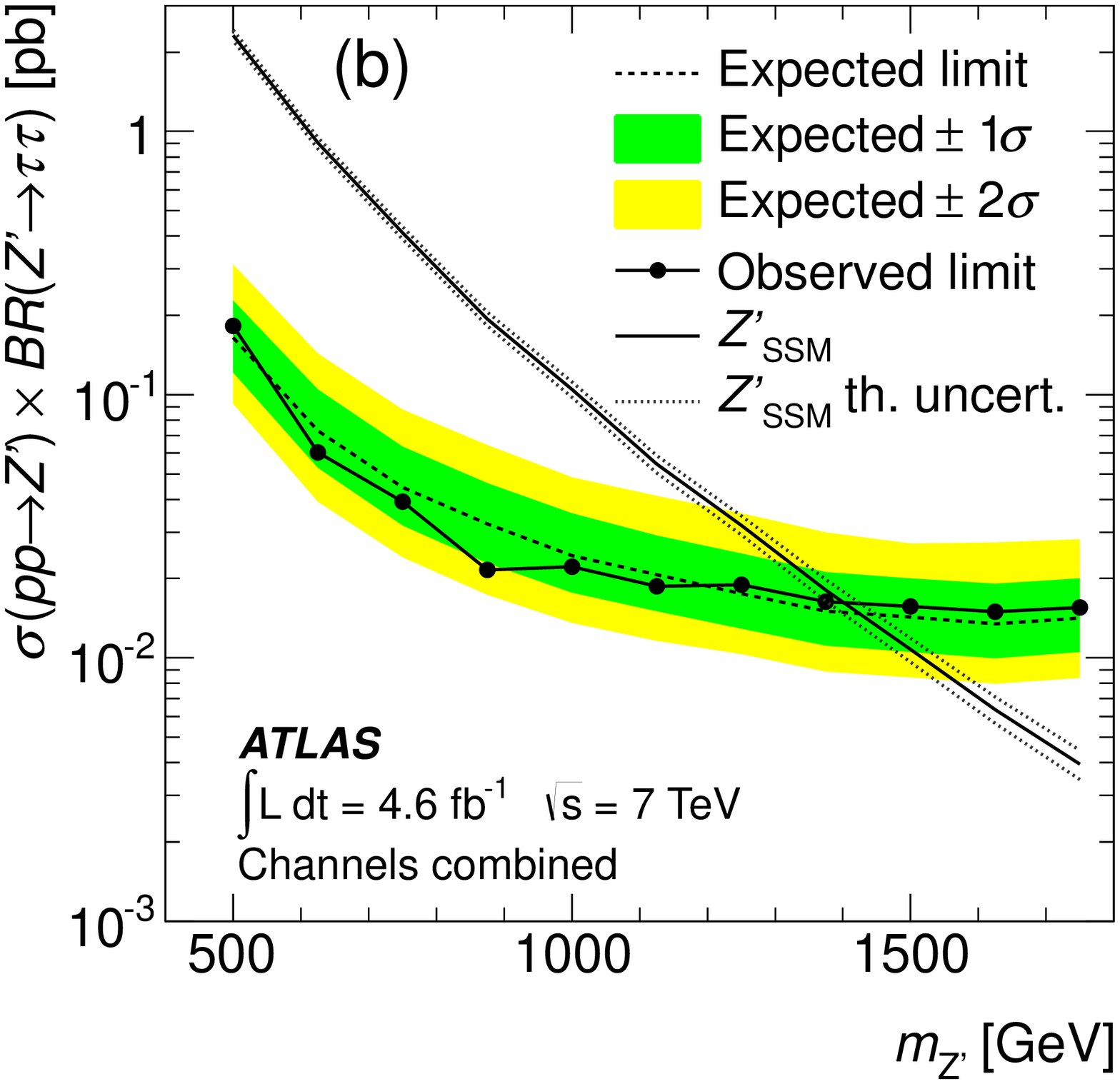}
    \caption{
        (a) The expected (dashed) and observed (solid) 95\% credibility upper limits on the
        cross section times \tptm{} branching fraction, in the $\hadhad{}$, 
        $\muhad{}$, $\ehad$ and $\emu$ channels and
        for the combination.  The expected \ZprimeSSM{} production cross section 
        and its corresponding theoretical uncertainty (dotted) are also included.  
        (b) The expected and observed limits for the
        combination including 1$\sigma$ and 2$\sigma$ uncertainty bands.  \ZprimeSSM{} 
        masses up to $1.40\TeV{}$ are excluded, in agreement with the expected limit of 
        $1.42\TeV$ in the absence of a signal.
    }
    \label{fig:results}
\end{figure*}

\section{Conclusion}
\label{sec:conclusion}

A search for high-mass ditau resonances has been performed using 4.6\,\ifb{}
of data collected with the ATLAS detector in $pp$ collisions at $\sqrt{s}=7\TeV$
at the LHC. 
The \hadhad{}, \muhad{}, \ehad{} and \emu{} channels are analysed.
The observed number of events in the high-transverse-mass region is consistent
with the SM expectation.  Limits are set on the cross section times branching
fraction for such resonances.  The resulting lower limit on the mass of a
\Zprime{} decaying to \tptm{} in the Sequential Standard Model is $1.40\TeV$ at
95\% credibility, in agreement with the expected limit of $1.42\TeV$
in the absence of a signal.




\section{Acknowledgements}

We thank CERN for the very successful operation of the LHC, as well as the
support staff from our institutions without whom ATLAS could not be
operated efficiently.

We acknowledge the support of ANPCyT, Argentina; YerPhI, Armenia; ARC,
Australia; BMWF and FWF, Austria; ANAS, Azerbaijan; SSTC, Belarus; CNPq and FAPESP,
Brazil; NSERC, NRC and CFI, Canada; CERN; CONICYT, Chile; CAS, MOST and NSFC,
China; COLCIENCIAS, Colombia; MSMT CR, MPO CR and VSC CR, Czech Republic;
DNRF, DNSRC and Lundbeck Foundation, Denmark; EPLANET, ERC and NSRF, European Union;
IN2P3-CNRS, CEA-DSM/IRFU, France; GNSF, Georgia; BMBF, DFG, HGF, MPG and AvH
Foundation, Germany; GSRT and NSRF, Greece; ISF, MINERVA, GIF, DIP and Benoziyo Center,
Israel; INFN, Italy; MEXT and JSPS, Japan; CNRST, Morocco; FOM and NWO,
Netherlands; BRF and RCN, Norway; MNiSW, Poland; GRICES and FCT, Portugal; MERYS
(MECTS), Romania; MES of Russia and ROSATOM, Russian Federation; JINR; MSTD,
Serbia; MSSR, Slovakia; ARRS and MVZT, Slovenia; DST/NRF, South Africa;
MICINN, Spain; SRC and Wallenberg Foundation, Sweden; SER, SNSF and Cantons of
Bern and Geneva, Switzerland; NSC, Taiwan; TAEK, Turkey; STFC, the Royal
Society and Leverhulme Trust, United Kingdom; DOE and NSF, United States of
America.

The crucial computing support from all WLCG partners is acknowledged
gratefully, in particular from CERN and the ATLAS Tier-1 facilities at
TRIUMF (Canada), NDGF (Denmark, Norway, Sweden), CC-IN2P3 (France),
KIT/GridKA (Germany), INFN-CNAF (Italy), NL-T1 (Netherlands), PIC (Spain),
ASGC (Taiwan), RAL (UK) and BNL (USA) and in the Tier-2 facilities
worldwide.

\newpage
\biboptions{compress}
\bibliographystyle{model1-num-names}

\onecolumn
\clearpage
\begin{flushleft}
{\Large The ATLAS Collaboration}

\bigskip

G.~Aad$^{\rm 48}$,
T.~Abajyan$^{\rm 21}$,
B.~Abbott$^{\rm 111}$,
J.~Abdallah$^{\rm 12}$,
S.~Abdel~Khalek$^{\rm 115}$,
A.A.~Abdelalim$^{\rm 49}$,
O.~Abdinov$^{\rm 11}$,
R.~Aben$^{\rm 105}$,
B.~Abi$^{\rm 112}$,
M.~Abolins$^{\rm 88}$
% ...
\bigskip

% ...
$^{11}$ Institute of Physics, Azerbaijan Academy of Sciences, Baku, Azerbaijan\\
$^{12}$ Institut de F\'{i}sica d'Altes Energies and Departament de F\'{i}sica de la Universitat Aut\`{o}noma de Barcelona and ICREA, Barcelona, Spain\\
% ...
$^{21}$ Physikalisches Institut, University of Bonn, Bonn, Germany\\
% ...
$^{48}$ Fakult\"{a}t f\"{u}r Mathematik und Physik, Albert-Ludwigs-Universit\"{a}t, Freiburg, Germany\\
$^{49}$ Section de Physique, Universit\'{e} de Gen\`{e}ve, Geneva, Switzerland\\
% ...
$^{88}$ Department of Physics and Astronomy, Michigan State University, East Lansing MI, United States of America\\
% ...
$^{105}$ Nikhef National Institute for Subatomic Physics and University of Amsterdam, Amsterdam, Netherlands\\
% ...
$^{111}$ Homer L. Dodge Department of Physics and Astronomy, University of Oklahoma, Norman OK, United States of America\\
$^{112}$ Department of Physics, Oklahoma State University, Stillwater OK, United States of America\\
% ...
$^{115}$ LAL, Universit\'{e} Paris-Sud and CNRS/IN2P3, Orsay, France\\
% ...
\end{flushleft}


\end{document}